\documentclass[%
reprint,%superscriptaddress,
groupedaddress,
longbibliography,
nobibnotes,
 amsmath,
 amssymb,
 aps,
 citeautoscript,
prb
]{revtex4-2}

\usepackage{graphicx}% Include figure files
\usepackage{multirow}
\usepackage{rotating}
\usepackage{ulem}
\usepackage{dcolumn}% Align table columns on decimal point
\usepackage{bm}% bold math
\usepackage{xcolor}
\usepackage{epstopdf}
\usepackage{epsfig}
\usepackage{booktabs}
\usepackage{array}
\usepackage{color,soul}
\usepackage{physics}
\usepackage{amsmath}
\newcolumntype{x}[1]{{\centering\hspace{0pt}}p{#1}}%
\graphicspath{ {Plots/} }
\newcommand{\RNum}[1]{\uppercase\expandafter{\romannumeral #1\relax}}
\newcommand{\comment}[1]{}
\begin{document}

%\preprint{APS/123-QED}

\title{Computational Discovery of Two-Dimensional Rare-Earth Iodides: Promising Ferrovalley Materials for Valleytronics}

\author{Abhishek Sharan}
\affiliation{Department of Physics, Khalifa University of Science and Technology, Abu Dhabi-127788, United Arab Emirates (UAE)}

\author{Stephan Lany}
\affiliation{National Renewable Energy Laboratory, Golden, CO 80401, USA}

\author{Nirpendra Singh}
\email{Nirpendra.Singh@ku.ac.ae}
\affiliation{Department of Physics, and Center for Catalysis and Separation (CeCaS), Khalifa University of Science and Technology, Abu Dhabi-127788, United Arab Emirates (UAE)}

\date{\today}% It is always \today, today,
             
\begin{abstract}
Two-dimensional \textit{Ferrovalley} materials with intrinsic valley polarization are rare but highly promising for valley-based nonvolatile random access memory and valley filter. Using Kinetically Limited Minimization (KLM), an unconstrained crystal structure prediction algorithm, and prototype sampling based on first-principles calculations, we have discovered 17 new \textit{Ferrovalley} materials (rare-earth iodides RI$_2$, where R is a rare-earth element belonging to Sc, Y, or La-Lu, and I is Iodine). The rare-earth iodides are layered and demonstrate 2H, 1T, or 1T$_d$ phase as the ground-state in bulk, analogous to transition metal dichalcogenides (TMDCs). The calculated exfoliation energy of monolayers is comparable to that of graphene and TMDCs, suggesting possible experimental synthesis. The monolayers in the 2H phase exhibit two-dimensional ferromagnetism due to unpaired electrons in $d$ and $f$ orbitals. Throughout the rare-earth series, $d$ bands show valley polarization at $K$ and $\overline{K}$ points in the Brillouin zone near the Fermi level. Due to strong magnetic exchange interaction and spin-orbit coupling, large intrinsic valley polarization in the range of 15-143 meV without external stimuli is observed, which can be tuned and enhanced by applying a biaxial strain. These valleys can selectively be probed and manipulated for information storage and processing, potentially offering superior performance beyond conventional electronics and spintronics. We further show that the 2H ferromagnetic phase of RI$_2$ monolayers possesses non-zero Berry curvature and exhibits the valley Hall effect with considerable anomalous Hall conductivity. Our work will incite exploratory synthesis of the predicted Ferrovalley materials and their application in valleytronics and beyond. 

\end{abstract}

%\pacs{Valid PACS appear here}% PACS, the Physics and Astronomy
                             % Classification Scheme.
%\keywords{Suggested keywords}%Use showkeys class option if keyword
                              %display desired
\maketitle
\clearpage
\section{Introduction}

Two-dimensional (2D) magnetism was a massive challenge until it was successfully demonstrated in layered Cr$_2$Ge$_2$Te$_6$ \cite{Gong2017} and CrI$_3$ \cite{Huang2017}. Magnetism combined with electronic and optical properties can lead to novel gateways for exploring magneto-electrical, magneto-optical, and valley properties of materials \cite{Lan2020, Burch2018, Wang2018, Deng2018, Klein2018}. In the last few years, 2D magnets have been used in novel magnetic devices to enhance the performance of existing devices, such as development of novel magnetic tunnel junctions (Fe$_3$GeTe$_2$/hBN/Fe$_3$GeTe$_2$) \cite{Wang2018}, gate tunable Curie temperature in Fe$_3$GeTe$_2$ over the extensive range 100-300 K \cite{Deng2018}, and giant tunnel magnetoresistance in CrI$_3$ based heterostructure \cite{Klein2018} to name a few. Certain 2D magnetic materials exhibit an extra valley degree of freedom for electrons in addition to charge and spin, which can be leveraged for memory devices leading to faster and more efficient logic systems and next-generation storage devices beyond conventional electronics and spintronics. Valley polarization is central to successfully controlling and manipulating valley degree of freedom, which requires breaking time-reversal and inversion symmetries. Since it was first observed and exploited in graphene \cite{Rycerz2007}, several approaches, such as optical pumping \cite{Zeng2012}, magnetic proximity effect \cite{Subhan2020, Zhao2017}, magnetic doping \cite{Singh2017}, and external magnetic \cite{MacNeill2015} and electric field \cite{Khan2021}  have been used to achieve valley polarization in transition metal dichalcogenides (TMDCs). However, these methods involve additional external complexity, such as strong interaction with magnetic substrates, dynamical processes for optical pumping, and external magnetic field, which is cumbersome for robust manipulation of valleys in device architecture.

2D \textit{Ferrovalley} materials possess exciting properties like 2D magnetism and intrinsic valley polarization and can be used in valley-based devices. These materials allow control and manipulation of valleys without any external perturbation. Till date only a handful of \textit{Ferrovalley} materials have been reported in literature, LaI$_2$, PrI$_2$ \cite{Sharan2022}, VSe$_2$ \cite{Tong2016}, GeSe \cite{Shen2017}, VAgP$_2$Se$_6$ \cite{Song2018}, CeI$_2$ \cite{Sheng2022}, LaBr$_2$ \cite{Zhao2019}, LaBrI \cite{Jiang2021}, CeI$_2$ \cite{Sheng2022}, NbX$_2$ (X=S, Se) \cite{Zang2020}, Nb$_3$I$_8$ \cite{Peng2020}, TiVI$_6$ \cite{Du2020} and GdI$_2$ \cite{Cheng2021}. Many more promising candidate materials are yet to be explored, and the expansion of the family of \textit{Ferrovalley} materials will accelerate the research interest in fabricating valley-based electronic devices. Furthermore, newly discovered materials can be combined with other 2D materials to form novel heterostructures exhibiting long-range ferromagnetism, complex spin textures, optoelectronic properties, topological bands, and ferroelectricity. This work explores the chemically rich space of rare-earth iodide compounds for their \textit{Ferrovalley} properties and valleytronics applications. Using unconstrained ground-state crystal structure prediction algorithm Kinetically Limited Minimization (KLM) and prototype sampling approach combined with first-principles calculations, we computationally predict several new two-dimensional \textit{Ferrovalley} materials. Rare-earth iodides (in RI$_2$ stoichiometry, where R ranges from Sc, Y, or La-Lu) are stable layered materials with ground-state structure either in 2H, 1T, or 1T$_d$ phase, analogous to TMDCs. These materials are magnetic in bulk and monolayer form, except for YI$_2$, which is non-magnetic. The 2H phase in monolayer form exhibits valley polarization in the 15-143 meV range due to strong magnetic exchange interaction and spin-orbit coupling. In the next few sections, we discuss the crystal structure, relative phase stability, exfoliation energy, electronic, magnetic, and anomalous Hall properties of predicted 2D RI$_2$ monolayers. 

\section{Methods}
The Kinetically Limited Minimization (KLM), an unconstrained crystal structure prediction algorithm, which combines unbiased random sampling and sequential atomic perturbations based on ground-state energy derived from the first principle calculations, is employed to predict the crystal structure of RI$_2$ materials. The KLM method has successfully been employed for the discovery of novel ternary nitrides \cite{Sun2019, Arca2018}, ternary oxynitrides \cite{Sharan2021}, and recently 2-D materials \cite{Sharan2022, Lany2021}. The crystal structure search is performed over 100-150 independent seed structures with varying formula units from 1 to 4. The initial random seed structures and subsequent perturbations enforce the minimum interatomic distance between different atomic species, significantly reducing the configurational space and prohibiting the inclusion of unphysical structures. The list of minimum inter-atomic distances used in the crystal structure search is included in the Supplementary Information (SI). The KLM approach demonstrates that the ground-state is either in 2H, 1T, and 1T$_d$ phase for all the RI$_2$ materials. Additionally, prototype structure sampling of RI$_2$ materials in 2H, 1T, and 1T$_d$ phases are performed to determine the low energy competing for meta-stable phases, which in some cases is not captured by the KLM method. The possible magnetic configurations, ferromagnetic (FM), anti-ferromagnetic (AFM), and non-magnetic (NM) are also taken into account for the ground-state crystal structure search. To address the intrinsic magnetism in these materials, a $2\times2\times1$ supercell for the 2H phase, $2\times2\times2$ supercell for the 1T phase, and $1\times2\times1$ supercell for the 1T$_d$ phase is constructed such that each unit cell contains two layers. For AFM configuration, the magnetic moment in two layers is initialized such that the nearest rare-earth atoms in two layers possess opposite magnetic moments. For FM and NM configurations, the respective unit cell is used to estimate the total energy.

The first-principles calculations are performed based on density functional theory with meta-GGA SCAN functional \cite{Sun2015} as implemented in the Vienna ab initio simulation package (VASP) \cite{Kresse1996a,Kresse1996b}. For faster convergence, generalized gradient approximation (GGA) of Perdew, Burke, and Ernzerhof (PBE) functional \cite{Perdew1996} is used for crystal structure sampling. Then further analysis is performed using SCAN functional on the determined structures. The interactions between ionic cores and valence electrons are described with projector-augmented-wave (PAW) potentials \cite{Blochl1994, Kresse1999}. The $3d$, $4s$ orbitals of Sc, $4s$, $4p$, $4d$, $5s$ orbitals of Y, and $4f$, $5s$, $5p$, $5d$, $6s$ orbitals of La-Lu rare earth elements are treated as valence electrons. The van der Waal corrections are incorporated using the DFT-D3 method \cite{Grimme2011}. Energy cutoff of 520 eV for plane-wave basis set expansion and $\Gamma$ centered $k$-mesh of $11\times11\times3$ for bulk and $11\times11\times1$ for monolayers for 2H and 1T phases, and $7\times12\times2$ for bulk and $7\times12\times1$ for monolayer for 1T$_d$ phase is used for integration over Brillouin zone. The atomic positions and lattice constant are optimized until forces are less than 0.01 eV/ \AA\ on each atom and total energy converged to less than 10$^{-6}$ eV. A vacuum of 20 \AA\ is used along the out-of-plane direction for monolayers to avoid interactions with its periodic images. The phonon band structure is computed using the density functional perturbation theory \cite{Gonze1997} as implemented in Phonopy code \cite{Togo2015}. The Berry curvature and anomalous Hall conductivity are calculated with WannierTools package \cite{Wu2018} with tight binding Hamiltonian extracted from WANNIER90 \cite{Mostofi2008}.

\section{Results}
\subsection{Crystal Structure}
Unconstrained crystal structure search taking into account various possible magnetic configurations, is executed for all possible rare-earth iodides with the general formula RI$_2$, where R is the rare-earth element from Sc, Y, or La-Lu and I is Iodine. We find that rare-earth iodides in RI$_2$ stoichiometry are layered materials either in trigonal prismatic (2H) phase belonging to $P6_3/mmc$ space group, octahedral (1T) phase belonging to $P\overline{3}m1$ space group, or distorted octahedral  (1T$_d$) phase belonging to  $Pnm2_1$ space group, as shown in Fig \ref{structure} (a), (b) and (c), respectively, analogous to TMDCs. Additionally, prototype structure sampling of RI$_2$ in 2H, 1T, and 1T$_d$ phase shows the relative stability of the polytypes (see Fig \ref{structure}(d)). The 2H and 1T phases are composed of a hexagonal lattice with 2 and 1 formula unit per unit cell, respectively, as shown by dashed lines in Fig \ref{structure}(a) and (b). 1T$_d$ phase can be generated from 1T phase by reconstruction in a 2$\times$1 orthorhombic cell and is composed of 4 formula units per unit cell. The reconstruction originated from a dimerized line of rare-earth atoms that leads to a lower energy structure, driven by distortion due to the Peierls transition mechanism \cite{Guyot1967, Lee1973}. Similar distortion leading to the stability of the 1T$_d$ phase in TMDCs has also been observed \cite{Besse2018, Keum2015}. All three phases are layered structures with van der Waals interactions between two subsequent layers. The equilibrium lattice parameters for each of the RI$_2$ in the three phases are determined and listed in Table \ref{table1}. The lattice parameters lie in a very narrow range throughout the series, (in-plane $a$ = 3.93 \AA\ - 4.60 \AA\, and out-of-plane per layer $c$ = 6.86 \AA\ - 8.18 \AA\ ), owing to similar atomic radius of the rare-earth elements. The distortion of the 1T$_d$ phase is not locally stable for all materials. In the case of PrI$_2$, NdI$_2$, SmI$_2$, EuI$_2$, DyI$_2$, ErI$_2$ and TmI$_2$, the initial 1T$_d$ structure relaxes into the 1T structure, suggesting that the 1T$_d$ phase is not even meta-stable for these materials. 

\subsection{Phase Stability}
\label{phst}
\begin{figure*}
\centering
\includegraphics[width=14 cm]{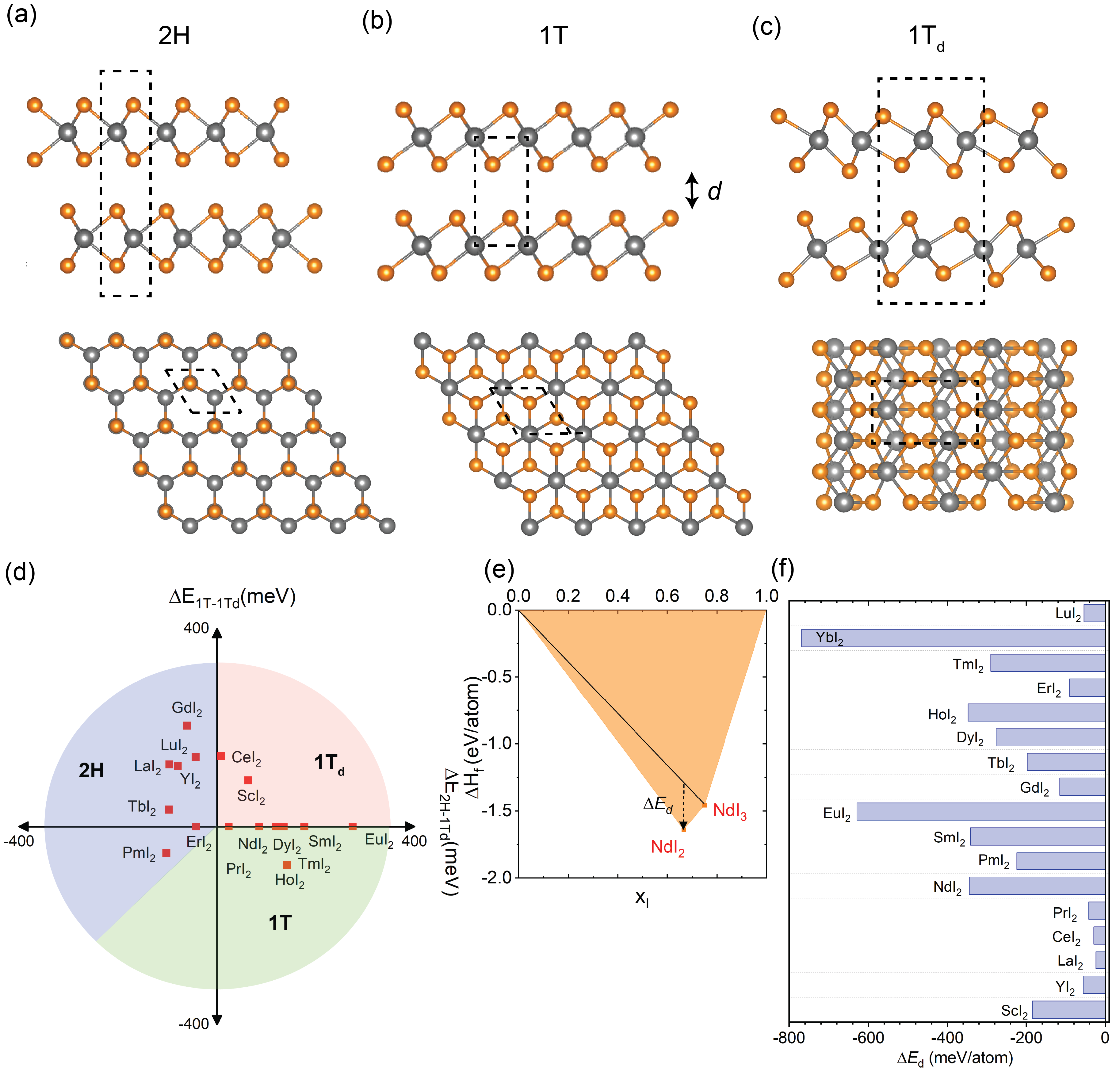}
\caption{Lateral and top view of (a) trigonal prismatic (2H), (b) octahedral (1T), and (c) distorted octahedral  (1T$_d$) phases of RI$_2$ compound. The unit cell is shown with a dashed line. (d) shows the relative stability of three phases for different RI$_2$ bulk compounds. $\Delta E_\mathrm{{2H-1T_d}}$ and $\Delta E_\mathrm{{1T-1T_d}}$ are plotted on the $x$ and $y$ axis respectively. The stability region for each phase is highlighted. (e) convex hull diagram of Nd-I system schematically showing the definition of decomposition energy, $\Delta E\mathrm{_d}$. (f) shows $\Delta E\mathrm{_d}$ in meV per atom, for different RI$_2$ bulk compounds.}
\label{structure}
\end{figure*}

Relative phase stability of RI$_2$ bulk compounds is shown in Fig \ref{structure}(d), where the phase stability is plotted in the space of $\mathrm{\Delta E_{2H-1T_d}}$ on $x$-axis and $\mathrm{\Delta E_{1T-1T_d}}$ on $y$-axis. The possible magnetic configurations, FM, AFM, and NM, are also taken into account for each RI$_2$ compound and the magnetic ground-state of each case is given in Table \ref{table1}. Out of the 17 RI$_2$ compounds, ScI$_2$, CeI$_2$ and YbI$_2$ have 1T$_d$ as the ground-state, while 2H and 1T phases are meta-stable. The lowest energy structure of YI$_2$, LaI$_2$, PmI$_2$, GdI$_2$, TbI$_2$, ErI$_2$ and LuI$_2$ is 2H phase, while 1T and 1T$_d$ are meta-stable phases except for ErI$_2$ where 1T$_d$ phase is unstable. For the case of PrI$_2$, NdI$_2$, SmI$_2$, EuI$_2$, DyI$_2$, HoI$_2$, and TmI$_2$ the ground-state is 1T phase while 2H phase makes meta-stable phase and 1T$_d$ is meta-stable only for HoI$_2$. For PrI$_2$, NdI$_2$, SmI$_2$, EuI$_2$, DyI$_2$, ErI$_2$, and TmI$_2$ that lie on $x$-axis in phase stability plot, Fig \ref{structure}(d), the 1T$_d$ is unstable. The 1T$_d$ phase transforms into lower energy 1T phase upon structural relaxation, hence $\mathrm{\Delta E_{1T-1T_d}} = 0$. 

 All the bulk RI$_2$ compounds have FM as the ground-state for magnetic configuration, except PmI$_2$, DyI$_2$, HoI$_2$, and ErI$_2$, which exhibit AFM as the ground-state, and YbI$_2$ is NM due to filled $4f$ orbitals in Yb$^{+2}$.  The $\Delta E_\mathrm{{2H-1T_d}}$ and $\Delta E_\mathrm{{1T-1T_d}}$ in Table \ref{table1} show slight energy differences between different structural phases, which suggests that synthesis of metastable phases may be possible at finite temperatures under appropriate experimental conditions. The thermodynamic stability of RI$_2$ against their competing phases, including the RI$_3$ phase, is investigated using convex hull analysis and computing the decomposition energy ($\Delta E\mathrm{_d}$), shown in Fig \ref{structure}(e) and (f). The $\Delta E\mathrm{_d}$ is defined as the energy difference of the lowest energy RI$_2$ phase from the convex hull formed by all the known phases (including elemental phase) between respective R and I atoms, shown in Fig \ref{structure}(e) for Nd-I system. A negative $\Delta E\mathrm{_d}$ suggests that the energy of the lowest energy RI$_2$ phase falls below the convex hull, as for NdI$_2$ in Fig \ref{structure}(e), and that it is stable against decomposition into its competing phases and can be experimentally synthesized, which is the case for all the RI$_2$ compounds predicted in this work (see Fig \ref{structure}(f)). It is to be noted that a three-dimensional phase of EuI$_2$ (space group 62 $Pnma$), which is already known \cite{Krings2009}, is lower in energy than the 1T-FM phase by 32 meV per formula unit. Towards I rich conditions, the RI$_3$ phase is an important composition for competing phases, as illustrated in Fig \ref{structure}(e). The ground-state energy of all the known phases in different stoichiometries is computed for the convex hull analysis. RI$_3$ compounds are known for all the rare-earth elements except for Eu, Gd, and Yb. Therefore, we perform crystal structure sampling for EuI$_3$, GdI$_3$, and YbI$_3$, using the KLM approach. We find that all three materials are stable with layered crystal structures belonging to space groups 2 ($P\overline{1}$), 162 ($P\overline{3}1m$), and 12 ($C2/m$) for EuI$_3$, GdI$_3$ and YbI$_3$ respectively. The crystallographic information file for EuI$_3$, GdI$_3$, and YbI$_3$ structures are available in SI.
 
Rare-earth iodides in RI$_3$ stoichiometry are also stable as layered materials, where rare-earth elements are in a +3 oxidation state. However, However, this work shows that rare-earth elements in a low-valent +2 oxidation state are also stable. R$^{+2}$ has the electron configuration [Ar]$3d^14s^0$ for Sc$^{+2}$, [Kr]$4d^{1}5s^0$ for Y$^{+2}$, and [Xe]$4f^{0-14}5d^{0-1}6s^0$ for R$^{+2}$ where R is La-Lu. From the electron configuration, one can say that the rare-earth elements are also early transition elements with zero or one $d$ electron in the valence shell. As with other transition elements which exhibit a range of oxidation states, rare-earth elements in R$^{0}$, R$^{+1}$, R$^{+2}$ and R$^{+3}$ oxidation states are also expected to exist \cite{Woen2016, Nief2010}. Considering other transition metal ions with similar electron configurations, for instance, Hf$^{+3}$ and La$^{+2}$ have the same electron configuration of [Xe]$5d^16s^0$, Hf$^{+3}$ is known in several compounds \cite{Fryzuk1996}. Therefore, the stability of rare-earth elements in a less common +2 oxidation state is also expected, as shown in this work for RI$_2$ compounds.

\subsection{Exfoliation Energy}
\begin{figure*}
\centering
\includegraphics[width=18 cm]{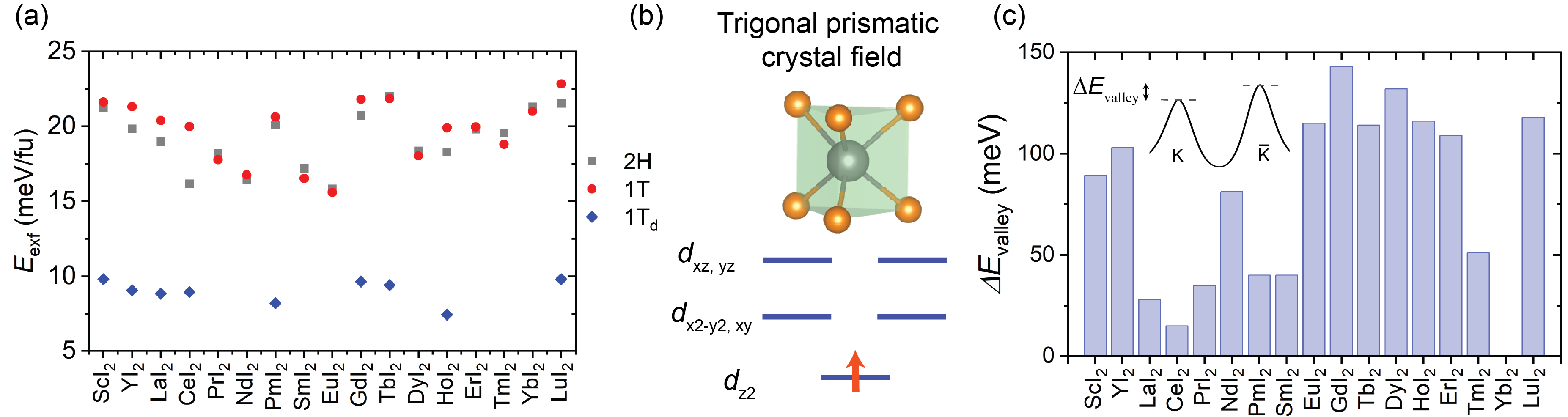}
\caption{(a) Exfoliation energy ($E_{\mathrm{exf}}$) of monolayer RI$_2$ in 2H, 1T and 1T$_d$ phase, (b) splitting of $d$ orbitals in trigonal prismatic crystal field of 2H phase, and (c) Valley polarization $\Delta E_{\mathrm{valley}}$ for different RI$_2$ monolayers in 2H-FM phase. }
\label{ML}
\end{figure*}

The exfoliation energy ($E_{\mathrm{exf}}$) of monolayers in different phases is calculated and listed in Table \ref{table1} and also shown in Fig \ref{ML} (a). The exfoliation energy is computed by the relation \cite{Jung2018},

\begin{equation*}
    E_{\mathrm{exf}} = (E_{\mathrm{mono}} - E_{\mathrm{bulk}})/A
\end{equation*} 

 where $E_{\mathrm{mono}}$ and $E_{\mathrm{bulk}}$ are the energy of monolayer and bulk per formula unit, respectively, and $A$ is the in-plane area of bulk. The calculated $E_{\mathrm{exf}}$ of monolayers are comparable to the exfoliation energy of graphene (12 meV/$\AA^2$) \cite{Liu2012} and MoS$_2$ (26 meV/$\AA^2$) \cite{Zhou2014}, advocating their mechanical exfoliation. The $E_{\mathrm{exf}}$ of 2H and 1T phases ranges between 16-23 meV/$\AA^2$, which is close to the exfoliation energies of TMDCs. The $E_{\mathrm{exf}}$ of distorted octahedral (1T$_d$) phase is significantly lower than that in 2H or 1T phase and ranges between 7-10 meV/$\AA^2$, which suggests that the monolayers in 1T$_d$ phase are weakly bound to each other and are easily exfoliable. The 1T$_d$ phase is derived from the 1T phase, where two neighbor metal atoms distort the octahedra around them, thereby increasing the out-of-plane lattice parameter due to Peierls distortion \cite{Guyot1967, Lee1973}, which effectively reduces the binding of adjacent layers, thereby decreasing the exfoliation energy.

\begin{table*}
\centering
\caption{Calculated lattice constants, ($a_0$, $b_0$ and $c_0$) in \AA\ and exfoliation energy ($E_{\mathrm{exf}}$) in $\mathrm{meV/\AA^2}$ for 2H, 1T and 1T$_d$ phases of RI$_2$ in bulk and monolayer form (for 2H and 1T phase $a_0$ = $b_0$). The energy difference between 2H and 1T$_d$ ($\Delta E_{\mathrm{2H-1T_d}}$), and between 1T and 1T$_d$ ($\Delta E_{\mathrm{1T-1T_d}}$) phase of bulk RI$_2$ is given in meV per formula unit (fu). The ground-state (GS) and corresponding magnetic configuration for bulk RI$_2$ (FM for ferromagnetic, AFM for anti-ferromagnetic, and NM for non-magnetic) is included. The decomposition energy, $\Delta E_\mathrm{d}$ in meV per atom, for each compound is also listed.}
\setlength{\tabcolsep}{4pt}
\setlength{\extrarowheight}{0pt}
\begin{tabular}{lc|ccc|ccc|cccc|c|c|c|c}
\hline 
                          &      &   \multicolumn{3}{c|}{2H Phase}  & \multicolumn{3}{c|}{1T Phase} & \multicolumn{4}{c|}{1T$_d$ Phase} & $\Delta E_\mathrm{{2H-1T_d}}$ & $\Delta E_\mathrm{{1T-1T_d}}$ & GS & $\Delta E_\mathrm{d}$ \\ 
                          &      & $a_0$ & $c_0$ & E$_{\mathrm{exf}}$  & $a_0$ & $c_0$ & E$_{\mathrm{exf}}$   & $a_0$ & $b_0$ & $c_0$ & E$_{\mathrm{exf}}$  & (meV/fu) & (meV/fu)& &(meV/at) \\ [0.01 cm]
                          \hline \hline
\multirow{2}{*}{ScI$_2$}  & bulk & 3.93  & 14.60   & &   4.01 & 7.05 &  &   4.02 & 6.90 & 14.35 & & 64 & 93 & \multirow{2}{*}{1T$_d$ - FM} & -185  \\
                          & ML   &   3.93   &     & 21   & 4.01 & & 22  & 4.02 & 6.90 & & 10 &    &   &   \\ 
                          \hline
\multirow{2}{*}{YI$_2$} & bulk & 4.06    & 15.10 &   & 4.12 & 7.31 &  &  4.09 & 7.18 & 15.03 &  & -79 & 123 & \multirow{2}{*}{2H - FM}     & -56      \\
                          & ML   &  4.06    &   & 20  & 4.12 & & 21  & 4.09 & 7.18 & & 9 &   &    &     \\ 
                          \hline
\multirow{2}{*}{LaI$_2$} & bulk & 4.26  & 15.35  &  & 4.26 & 7.50 &  & 4.25 & 7.55 & 15.27  & & -96 & 126 & \multirow{2}{*}{2H - FM} & -24 \\
                          & ML   &  4.26    & & 19   & 4.26 & & 19 &  4.25 & 7.55 &  & 9 &  &    &     \\ 
                          \hline
\multirow{2}{*}{CeI$_2$}  & bulk & 4.19  & 15.30  & & 4.12 & 7.64  & & 4.25 & 7.42 & 15.11  & & 8 & 143 & \multirow{2}{*}{1T$_d$ - FM} & -29 \\
                          & ML & 4.19 & & 16   &  4.12 & & 10  &   4.25 & 7.42 & &  9 &   &  &    \\ 
                          \hline
\multirow{2}{*}{PrI$_2$} & bulk & 4.28 & 15.14  & & 4.45 & 7.24 &  & \multicolumn{4}{c|}{unstable} & 24 & 0 & \multirow{2}{*}{1T - FM} & -42 \\
                          & ML &  4.28 & & 18 &  4.45 & & 18 &   &  &  &  &   & & \\ 
                          \hline
\multirow{2}{*}{NdI$_2$} & bulk & 4.35  & 14.93  & & 4.56 & 7.00  & & \multicolumn{4}{c|}{unstable}  & 119 & 0 & \multirow{2}{*}{1T - FM} & -345   \\
                          & ML   &  4.35 & & 16   & 4.56 & & 17    &    &  &   & &  & & \\ 
                          \hline
 \multirow{2}{*}{PmI$_2$} & bulk & 4.22  & 14.77  & & 4.36 & 7.22  & & 4.28 & 7.40 & 14.91  & & -102 & -53 & \multirow{2}{*}{2H - AFM} & -224 \\
                          & ML &  4.22 & & 19 &  4.36 & & 21  & 4.28 & 7.40 &  & 8 &  &   &  \\ 
                          \hline
\multirow{2}{*}{SmI$_2$} & bulk & 4.35    & 14.81     & & 4.57 & 6.94  &   & \multicolumn{4}{c|}{unstable} & 177 & 0 & \multirow{2}{*}{1T - FM}    & -342      \\
                          & ML   &  4.35    &      & 17 & 4.57 & & 17  &  &  & &  & &   &    \\ 
                          \hline
\multirow{2}{*}{EuI$_2$} & bulk & 4.43 & 14.78  & & 4.60 & 6.89  & & \multicolumn{4}{c|}{unstable} & 275 & 0 & \multirow{2}{*}{1T - FM} & -629 \\
                          & ML & 4.43 &  & 16  & 4.60 & & 16 &   &  & &  &  & & \\ 
                          \hline
\multirow{2}{*}{GdI$_2$} & bulk & 4.09 & 14.99   & & 4.06 & 7.45  & & 4.11 & 7.18 & 14.99  & & -60 & 204 & \multirow{2}{*}{2H - FM} & -116 \\
                          & ML &  4.09 & & 21 &  4.06 & & 22  & 4.11 & 7.18 & & 10  & & & \\ 
                          \hline
\multirow{2}{*}{TbI$_2$} & bulk & 4.05    & 15.01 & & 4.09 & 7.33 &    & 4.29 & 7.56 & 16.32 &  & -97 & 34 &    \multirow{2}{*}{2H - FM}   & -198   \\
                          & ML   &  4.05    &  & 22 & 4.09 & &  22  & 4.29 & 7.56 & & 8 &   &  &   \\ 
                          \hline
\multirow{2}{*}{DyI$_2$} & bulk & 4.22 & 14.81    & & 4.37 & 7.05  &  & \multicolumn{4}{c|}{unstable} & 86 & 0 & \multirow{2}{*}{1T - AFM} & -276 \\
                          & ML &  4.22 &    & 18  & 4.37 & & 19  &  &  & &  & &    &       \\ 
                          \hline
\multirow{2}{*}{HoI$_2$} & bulk & 4.25  & 14.73  & & 4.32 & 6.86  &  & 4.38 & 7.47 & 14.46  & & 142 & -77 & \multirow{2}{*}{1T - AFM} & -348 \\
                          & ML &  4.25 & & 17   & 4.32 & & 18  & 4.38 & 7.47 & & 7  & &  &   \\ 
                          \hline
\multirow{2}{*}{ErI$_2$} & bulk & 4.11 & 14.56  & & 4.22 & 6.99  & & \multicolumn{4}{c|}{unstable} & -42 & 0 & \multirow{2}{*}{2H - AFM} & -91 \\
                          & ML &  4.11 & & 23  & 4.22 & & 21  &  &  & &   & & & \\ 
                          \hline
\multirow{2}{*}{TmI$_2$} & bulk & 4.13 & 14.66  & & 4.26 & 7.09  & & \multicolumn{4}{c|}{unstable} & 135 & 0 & \multirow{2}{*}{1T - FM}  & -290 \\
                          & ML & 4.13 &   & 20  & 4.26 & & 19  &  &  & &  &  &     &  \\ 
                          \hline
\multirow{2}{*}{YbI$_2$} & bulk & 3.96    & 14.79    & & 4.02 & 7.20  &  & 3.99 & 7.00 & 14.85 & & 488 & 219  & \multirow{2}{*}{1T$_d$ - NM}  & -769 \\
                          & ML   &  3.96    &    &   21     & 4.02 & & 21 & 3.99 & 7.00 & & 8 &  & &    \\ 
                          \hline
\multirow{2}{*}{LuI$_2$} & bulk & 3.97 & 14.91  & & 4.02 & 7.16  & & 3.99 & 6.98 & 14.95  & & -43 & 141 & \multirow{2}{*}{2H - FM} & -54 \\
                          & ML & 3.97 & & 22  & 4.02 & & 23  & 3.99 & 6.98 & & 10  & & & \\          
                          \hline \hline
\end{tabular}
\label{table1}
\end{table*}

\subsection{Electronic Properties of 2H-monolayer}

 The electronic band structure of RI$_2$ monolayer (ML) in the 2H phase shows valley polarization and is interesting for valleytronics. Therefore, the discussion will focus mainly on the 2H-ML phase of RI$_2$ compounds unless otherwise stated. The 2H phase in ML form is the ground-state for YI$_2$, LaI$_2$, CeI$_2$, PmI$_2$, GdI$_2$, TbI$_2$, ErI$_2$, TmI$_2$ and LuI$_2$ MLs, while 1T phase of PrI$_2$, NdI$_2$, SmI$_2$, EuI$_2$, DyI$_2$, HoI$_2$ and YbI$_2$ MLs is the ground-state and 1T$_d$ phase of ScI$_2$ ML has the lowest energy. All the RI$_2$ MLs favor FM configuration, except for DyI$_2$ (AFM), ErI$_2$ (AFM), and YbI$_2$ (NM), as mentioned in Table \ref{table2}. For a few cases, the magnetic configuration differs from the respective bulk ground-state of RI$_2$ compounds, as discussed in section \ref{phst} and listed in Table \ref{table1}. The energy difference between the lowest energy phase and the 2H$-$FM configuration for MLs, $\Delta E_{\mathrm{2H, FM}}$, is shown for comparison in Table \ref{table2}. The energy difference between stable and metastable phases considering different magnetic configurations is comparable to room temperature for several cases, suggesting that, under suitable experimental conditions, the 2H phase in monolayer form can potentially be synthesized for all the RI$_2$ compounds. The dynamical stability of RI$_2$ MLs in the 2H phase is demonstrated by the absence of negative frequencies in phonon dispersion plots, shown in Fig S1.
 
\begin{figure*}
\centering
\includegraphics[width=16 cm]{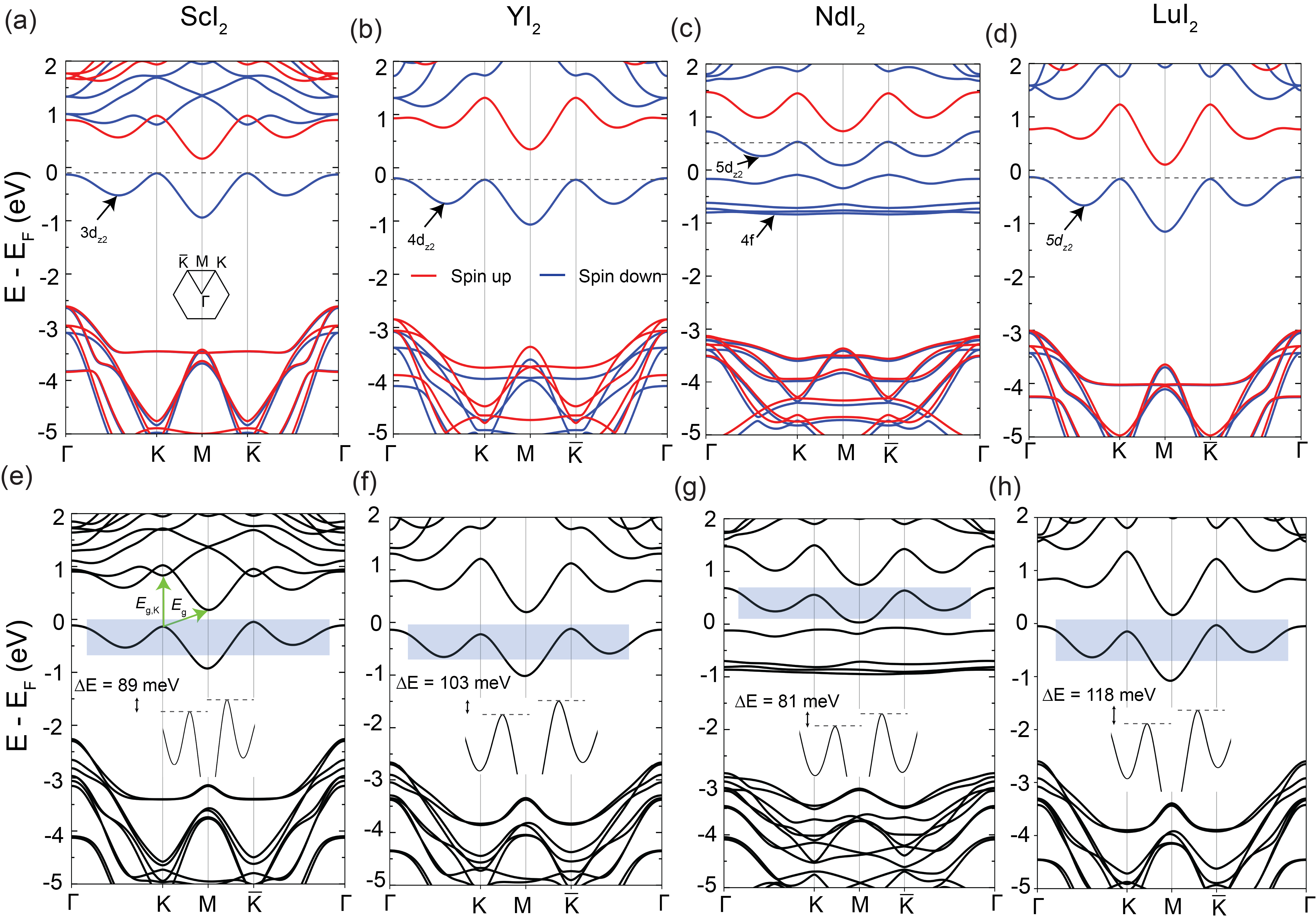}
\caption{Calculated band structure of (a) ScI$_2$, (b) YI$_2$, (c) NdI$_2$, and (d) LuI$_2$ monolayers without SOC in 2H-FM configuration. Spin up and spin down bands are plotted in red and blue colors, respectively. Valley degeneracy at $K$ and $\overline{K}$ points are indicated by a dashed line in each case. The inset in (a) shows the high symmetry lines in the Brillouin zone along which the bandstructures are plotted. The $3d_{z2}$, $4d_{z2}$, $5d_{z2}$ and $4f$ states for ScI$_2$, YI$_2$, NdI$_2$ and Lu$_2$ is also shown in (a)-(d). (e), (f), (g) and (h) shows the SOC band structure of ScI$_2$, YI$_2$, NdI$_2$, and LuI$_2$ in 2H-FM configuration. The broken valley degeneracy is highlighted in the zoomed in inset in each case. Valley polarization values are also mentioned. The $E_\mathrm{g}$ and $E_\mathrm{g,K}$ are schematically shown in (e).}
\label{bandstr}
\end{figure*}

 In the trigonal prismatic crystal field of 2H phase, each R atom is coordinated with 6 I atoms. The $d$ orbitals of R element split into three groups $a$($d_{z2}$), $e_1$($d_{x2-y2,xy}$) and $e_2$($d_{xz,yz}$) with increasing energy as shown in Fig \ref{ML}(b). The $3d_{z2}$ of Sc$^{+2}$ and $4d_{z2}$ of Y$^{+2}$ are singly occupied in spin-up channel giving rise to a magnetic moment of 1 $\mu_B$ per formula unit with FM ordering. For other RI$_2$ monolayers, the partial occupancy of $4f$ and $5d$ orbitals give rise to magnetism in these monolayers with magnetic moments given in Table \ref{table2}. Depending on the number of electrons in $4f$ orbitals, the $5d_{z2}$ orbitals is either occupied or unoccupied but lies near the Fermi energy ($E_F$).  Irrespective of the occupation of $d_{z2}$ orbitals, the dispersion of $d_{z2}$ state near the $E_F$ gives rise to the valleys at $K$ and $\overline{K}$ points in the Brillouin zone. The electronic band structure of ScI$_2$, YI$_2$, NdI$_2$, and LuI$_2$ monolayers in 2H$-$FM configuration, with and without spin-orbit coupling (SOC) is shown in Fig \ref{bandstr}. With SOC, ScI$_2$, YI$_2$, NdI$_2$, and LuI$_2$ monolayers in 2H phase are FM semiconductors with band gap of 0.23 eV, 0.32 eV, 0.10 eV, and 0.19 eV, and magnetic moments of 0.7 $\mu_B$, 0.5 $\mu_B$, 3.5 $\mu_B$ and 0.5 $\mu_B$ per formula unit, respectively. The band structure of other RI$_2$ MLs in 2H phase are shown in Fig S3-S6. LaI$_2$, CeI$_2$, PmI$_2$, SmI$_2$, EuI$_2$, GdI$_2$ and LuI$_2$ monolayers are FM semiconductors with small band gap varying from 0.08 eV (LaI$_2$) to 0.77 eV (EuI$_2$). PrI$_2$, TbI$_2$, DyI$_2$, HoI$_2$, ErI$_2$ and TmI$_2$ are FM metallic due to partial occupation of $4f$ and $5d$ orbitals. YbI$_2$ is a NM semiconductor due to fully filled $4f$ orbitals of Yb$^{+2}$. The indirect band gap $E_g$ of all the RI$_2$ MLs in 2H phase is given in Table \ref{table2}. The direct band gap at $K$, $E_{g,K}$ is also given in Table \ref{table2} and schematically shown in Fig \ref{bandstr}(e). Without SOC, the valleys at $K$ and $\overline{K}$ are degenerate, as highlighted by dashed lines in Fig \ref{bandstr} (a)-(d). The SOC lifts the degeneracy, leading to a large spontaneous valley polarization ($\Delta E_{\mathrm{valley}}$) of 89 meV, 103 meV, 81 meV and 118 meV in ScI$_2$, YI$_2$, NdI$_2$, and LuI$_2$ MLs respectively, also shown in Fig \ref{bandstr}(e)-(f). A large valley polarization is also seen in other RI$_2$ MLs varying in the range of 15-143 meV, as shown in their electronic band structures Fig S3-S6, listed in Table \ref{table2} and shown Fig \ref{ML}(c). The largest value of 143 meV for $\Delta E_{\mathrm{valley}}$ is for GdI$_2$ monolayer which also possess largest magnetic moment of 7.4 $\mu_B$ in FM configuration. Although the indirect band gap in these MLs is small, there is a large separation between the valence band maximum at $K$ and conduction band minimum at $M$, ensuring the robustness of valleys against scattering by long-wavelength phonons. Therefore, these materials exhibit intrinsic magnetism in two dimensions and show large spontaneous valley polarization without any external stimuli. Additionally, these valleys are spin-polarized and coupled to the magnetic ordering, i.e., the polarity of valleys can be reversed by changing the direction of intrinsic magnetism.

In certain cases, due to the relative ordering of $5d_{z2}$ and $4f$ orbitals, the $5d_{z2}$ orbitals lie slightly away from $E_F$. Such as, in case of PrI$_2$ monolayer the valley band is $\sim$ 1 eV below the $E_F$ (Fig S3(c) and (g)), while for NdI$_2$ monolayer the valley band is $\sim$ 0.5 eV above $E_F$ (Fig \ref{bandstr}(c) and (g)). The $E_F$ needs to be tuned to access the valley states in such cases. Such $E_F$ tuning has been successfully achieved in Bi$_2$Se$_3$, and Ca / Sn doped Sb$_2$Te$_2$Se \cite{Wang2011}, without modifying the states near the $E_F$. The valleys in SmI$_2$, TbI$_2$, DyI$_2$ and TmI$_2$ MLs are difficult to access due to the presence of $4f$ states near the $E_F$ (Fig S3-S6). Nevertheless, we have included the valley polarization in Table \ref{table2} and Fig \ref{ML}(c) for all the RI$_2$ monolayers in 2H-FM configuration. The valley polarization can further be enhanced and tuned by the application of biaxial strain \cite{Sharan2022}. The spin-polarized valleys give an extra degree of freedom for the electrons to possess, which can potentially be utilized for enhanced functionalities of next-generation devices. The traditional valley materials require an external source such as magnetic proximity effect \cite{Qi2015}, optical pumping \cite{Zeng2012}, or magnetic field \cite{MacNeill2015} to achieve valley polarization. In contrast, RI$_2$ \textit{Ferrovalley} materials eliminate the need for any external effect and are easier to implement in device architecture, making them emerging materials for next-generation nonvolatile memory devices \cite{Sakamoto2013, Ahammed2022}.

It is worth mentioning that the value of valley polarization and position of $4f$ and $5d$ states of a rare-earth element is dependent on the exchange-correlation functional. For instance, when using Generalized Gradient Approximation (GGA) functional of Perdew-Burke-Ernzerhof (PBE) \cite{Perdew1996}, valley polarization of 10 meV and 55 meV is computed for LaI$_2$ and PrI$_2$ MLs respectively, compared to 28 meV and 35 meV using meta-GGA SCAN functional \cite{Sun2015}. The Hubbard potential parameter ($U$) correctly describes crystal lattice, and electronic properties of rare-earth or transition metal  \cite{Li2020}. However, the $U$ value is often used as a fitting parameter, and its value is ambiguous in describing the properties of a novel system. To avoid such ambiguity, we have reported the results based on SCAN functional, which better describes the properties of correlated systems than the PBE functional \cite{Kitchaev2016, Chakraborty2018}. However, it is critical to mention here that although the quantitative description of valley polarization reported in this work may lie within a certain energy window, due to the limitation of exchange-correlation functionals in the accurate description of $4f$ and $5d$ state of rare-earth elements, a qualitative description of the discovery of magnetism in two dimensions and \textit{Ferrovalley} property in these materials and trends in valley polarization across the rare-earth series certainly holds true.

\begin{table*}
\centering
\caption{Valley polarization ($\Delta E_{\mathrm{valley}}$), global band gap ($E_\mathrm{g}$), direct band gap at $K$ point ($E_{\mathrm{g,K}}$), and magnetic moment ($\mu_B$) per formula unit for RI$_2$ monolayers in 2H-FM configuration. The ground-state ($GS_{\mathrm{ML}}$) of each monolayer is given and ground-state energy with respect to 2H-FM configuration ($\Delta E_{\mathrm{GS-2H,FM}}$) in meV per formula unit is also shown.}
\setlength{\tabcolsep}{4pt}
\setlength{\extrarowheight}{4pt}
\begin{tabular}{c|c|c|c|c|c|c} 
\hline
     & $\Delta E_{\mathrm{valley(2H,FM)}}$ (meV) & $E_\mathrm{g(2H,FM)}$ (eV) & $E_{\mathrm{g,K(2H,FM)}}$ (eV) & $(\mu_B$ per fu)$_\mathrm{{2H,FM}}$ & $GS_{\mathrm{ML}}$ & $\Delta E_{\mathrm{GS-2H,FM}}$ (meV/fu) \\ 
\hline
\hline
ScI$_2$ & 89                    & 0.23          & 0.86 & 0.7   & 1T$_d$ - FM & -76        \\
YI$_2$  & 103                   & 0.32          & 1.24 & 0.5  & 2H - FM &  0       \\
LaI$_2$ & 28                    & 0.08          & 0.74 & 0.4    & 2H - FM & 0       \\
CeI$_2$ & 15                    & 0.39          & 0.50 & 1.4    & 2H - FM & 0       \\
PrI$_2$ & 35                    & metallic      & 0.76 & 2.5    & 1T - FM & -8       \\
NdI$_2$ & 81                    & 0.10          & 0.79 & 3.5    & 1T - FM & -86       \\
PmI$_2$ & 68                    & 0.22          & 1.00 & 4.5    & 2H - FM & 0       \\
SmI$_2$ & 40                    & 0.18          & 0.14 & 5.5    & 1T - FM & -160       \\
EuI$_2$ & 115                   & 0.77          & 0.60 & 6.8    & 1T - FM & -258       \\
GdI$_2$ & 143                   & 0.60          & 0.58 & 7.4    & 2H - FM & 0       \\
TbI$_2$ & 114                   & metallic      & 0.05 & 6.1     & 2H - FM & 0       \\
DyI$_2$ & 132                   & metallic      & 0.15 & 4.5   & 1T - AFM & -70        \\
HoI$_2$ & 116                   & metallic      & 0.69 & 3.4    & 1T - FM & -184       \\
ErI$_2$ & 109                   & metallic      & 0.86 & 2.7    & 2H - AFM & -668        \\
TmI$_2$ & 51                    & metallic      & 0.39 & 1.2    & 2H - FM & 0        \\
YbI$_2$ & 0                     & 0.31          & 0.00    & 0.0 & 1T - NM & -265          \\
LuI$_2$ & 118                   & 0.19          & 1.25 & 0.5    & 2H - FM & 0        \\
\hline
\hline
\end{tabular}
\label{table2}
\end{table*}

\subsection{Valley Hall Effect}
In 2D hexagonal magnetic systems with broken inversion and time-reversal symmetries, the charge carriers possess non-zero Berry curvature $\Omega_z(k)$ at $K$ and $\overline{K}$ points in the Brillouin zone, leading to valley contrasting features in electron transport, known as the valley Hall effect \cite{Xiao2012, Rycerz2007, Mak2014, Pan2014}. The impact of an external electric field on electron transport is investigated by calculating the out of plane ($z$) component of the Berry curvature using the Kubo formula \cite{Thouless1982}:

\begin{equation}
\label{berry_cur}
\Omega_z (k)= -\sum_{m} \sum_{m \neq n} f(m) \frac{\mathrm{2Im} \bra{\phi_{mk}} \nu_x \ket{\phi_{nk}} \bra{\phi_{nk}} \nu_y \ket{\phi_{mk}}}{(E_m(k) - E_n(k))^2},
\end{equation}

where $m$ and $n$ are the band index for the occupied orbitals, $f(m)$ is the Fermi-Dirac distribution function for the $m^{th}$ band, $\nu_x(y)$ is the velocity operator of electrons along $x(y)$ directions, and $\ket{\phi_{mk}}$ and $\ket{\phi_{nk}}$ are the periodic part of the Bloch wavefunctions with eigenvalues $E_m(k)$ and $E_n(k)$ respectively. The valley contrasting features in $\Omega_z(k)$ for ScI$_2$, YI$_2$, NdI$_2$, and LuI$_2$ 2H MLs is shown in Fig \ref{berry_cur}. The peaks of $\Omega_z(k)$ that are opposite in sign and unequal in magnitude can be seen at $K$ and $\overline{K}$ points. Under the effect of an external in-plane electric field, $E$, the electrons in these monolayers acquire velocity along transverse direction, $v_{\bot} \sim E_{\parallel}$ $\times$ $\Omega_z(k)$, giving rise to anomalous Hall conductivity ($\sigma_{xy}$). The $\sigma_{xy}$ is a function of the $E_F$ and is computed as follows:

\begin{equation}
    \label{ahc}
    \sigma_{xy} = \frac{e^2}{\hbar} \int_{BZ} \frac{d^2k}{(2\pi)^2}f(k)\Omega_z(k)
\end{equation}

where $f(k)$ is the Fermi-Dirac function as a function of reciprocal lattice wave vector $k$.

The $\sigma_{xy}$ of ScI$_2$, YI$_2$, NdI$_2$, and LuI$_2$ monolayers as a function of the $E_F$ is shown in Fig \ref{berry_cur}(e). The zero level means that the $d^{\downarrow}$ bands up to the valley (also highlighted in Fig \ref{bandstr}) are occupied, and $d^{\uparrow}$ bands are unoccupied. $\sigma_{xy}$ attains a large value when the $E_F$ is negative and crosses the valleys. The similarity in the variation of $\sigma_{xy}$ in ScI$_2$, YI$_2$, and LuI$_2$ monolayers is a direct consequence of the similar nature of $d_{z2}$ orbitals near the $E_F$. $\sigma_{xy}$ attains large value near the near the $E_F$ in NdI$_2$ than ScI$_2$, YI$_2$ or LuI$_2$ due to the contribution of $4f$ states near the $E_F$ to $\sigma_{xy}$. The variation of $\sigma_{xy}$ of all other RI$_2$ monolayers is shown in Fig. S7 of SI. It is to be noted that valleys at the $\Gamma$ point do not contribute to $\Omega_z(k)$ or $\sigma_{xy}$, and therefore, these charge carriers will be undeflected in the transverse direction in a Hall setup. Due to significant anomalous Hall conductivity in these monolayers, the charge carriers along valleys can selectively be controlled and manipulated for valleytronic applications. 

\begin{figure*}
\centering
\includegraphics[width=16 cm]{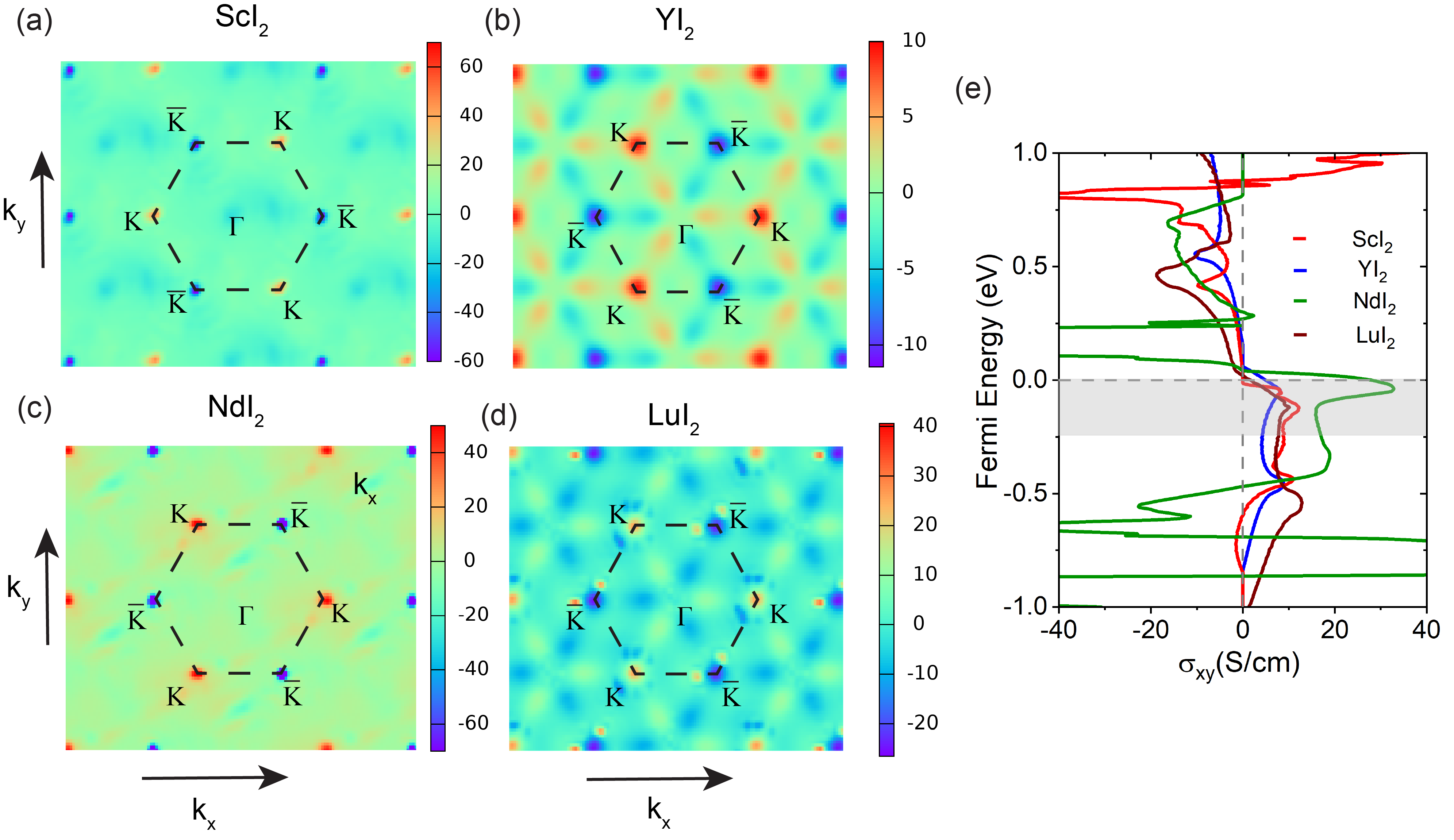}
\caption{Distribution of out of plane ($z$) component of Berry curvature ($\Omega_z$) for 2H-FM phase of (a) ScI$_2$, (b) YI$_2$, (c) NdI$_2$, and (d) LuI$_2$ monolayers within the reciprocal space. First Brillouin zone and high symmetry points $K$ and $\overline{K}$ are highlighted. (e) shows the variation of Anomalous Hall conductivity, $\sigma_{xy}$ as a function of $E_F$. The region with $E_F$ from 0 to -0.25 eV is highlighted.}
\label{berry_cur}
\end{figure*}

\section{Conclusions}
The unconstrained ground-state crystal structure prediction algorithm and prototype sampling have been employed to discover new 2D rare earth iodides (RI$_2$ materials where R is a rare-earth element from Sc, Y, and La-Lu, and I is Iodine). All the bulk materials are layered and either have trigonal prismatic (2H-$P6_3/mmc$ space group), octahedral (1T- $P\overline{3}m1$ space group), or distorted octahedral (1T$_\mathrm{d}$-$Pnm2_1$ space group) phase as the lowest energy state. The thermodynamic stability analysis shows that all the compounds are thermodynamically stable. The exfoliation energy of monolayers is comparable to graphene and TMDCs, suggesting possible mechanical exfoliation in the experiment. ScI$_2$, YI$_2$, LaI$_2$, CeI$_2$, NdI$_2$, PmI$_2$, SmI$_2$, EuI$_2$, GdI$_2$ and LuI$_2$ monolayers in 2H phase are FM semiconductors with small band gap varying from 0.08 eV (LaI$_2$) to 0.77 eV (EuI$_2$). PrI$_2$, TbI$_2$, DyI$_2$, HoI$_2$, ErI$_2$ and TmI$_2$ monolayers in 2H phase are FM metals, while YbI$_2$ is a NM semiconductor. RI$_2$ monolayers exhibit large intrinsic valley polarization in the range of 15-143 meV in the 2H-FM phase, which can further be tuned and enhanced on the application of biaxial strain. These materials are novel \textit{Ferrovalley} materials, where electrons possess the additional valley degree of freedom along with charge and spin. Beyond conventional electronics and spintronics, these materials are emerging for next-generation electronic devices for information processing and data storage. We further show that these monolayers exhibit the valley Hall effect with considerable anomalous Hall conductivity up to 30 S/cm with proper tuning of Fermi level. Because of the rapid development of 2D materials and their applications, discovering new 2D materials with magnetism and large intrinsic valley polarization without external stimuli is significant for valleytronics application. We believe that incite exploratory synthesis, and the related applications of 2D rare earth iodides in valleytronics will be highly sought after.

\section*{Supplementary Material}
See supplementary material for the structure sampling parameters, phonon dispersion, details on convex hull analysis, electronic bandstructres of RI$_2$ materials and crystallographic information file for EuI$_3$, GdI$_3$, and YbI$_3$ structures.
 Table S1: Minimum inter-atomic distances $d_{min}$ used in KLM structure sampling.
 Figure S1: Phonon dispersion of RI$_2$ monolayers in 2H phase.
 Figure S2: Convex Hull diagrams.
 Figure S3-S6: Electronic bandstructures of RI$_2$ monolayers in 2H-FM phase.
 Figure S7: Anomalous Hall conductivity as a function of the Fermi level for RI$_2$ monolayers in 2H-FM phase.
\section*{Acknowledgements} A. S. and N. S. acknowledge the financial support from Khalifa University of Science and Technology under the startup grant FSU-2020-11/2020. 
%\section*{Data Availability Statement}
\clearpage
\bibliography{references}
\end{document}